# Atomic-scale insights on grain boundary segregation in a cathode battery material

Inger-Emma Nylund[a], Elise R. Eilertsen[a], Constantinos A. Hatzoglou[a], Kaja Eggen Aune[b], Ruben Bjørge[b,c], Antonius T. J. van Helvoort[b], Ann Mari Svensson[a], Paraskevas Kontis[a,*]

[a] Department of Materials Science and Engineering, NTNU Norwegian University of Science and Technology, 7491 Trondheim, Norway
[b] Department of Physcis, NTNU Norwegian University of Science and Technology, 7491 Trondheim, Norway
[c] SINTEF Industry, 7465 Trondheim, Norway

*corresponding author: paraskevas.kontis@ntnu.no

## Abstract

The grain boundary segregation of the Co-free pristine spinel $LiNi_{0.5}Mn_{1.5}O_4$ cathode material has been studied by transmission electron microscopy and atom probe tomography. The segregation of Mn and the depletion of Ni at grain boundaries was observed by electron energy loss spectroscopy. These observations were also confirmed by atom probe tomography at other grain boundaries, which also revealed segregation of O and depletion of Li at grain boundaries. In addition, both methods revealed the occurrence of grain boundary segregation of Na, which is an impurity. Finally, segregation of O and Mn and a depletion of Li are also observed at dislocations. This observation has the potential to provide further support for the segregation behavior at the grain boundaries of this cathode material. These near-atomic-scale observations provide new insights that can be used to improve the synthesis and efficiency of Li-ion battery materials.

**Keywords:** grain boundary, segregation, cathode, transmission electron microscopy, atom probe tomography

The high-voltage spinel $LiNi_{0.5}Mn_{1.5}O_4$ (LNMO) is a promising Co-free cathode material for the Li-ion battery (LIB) systems. Given its high energy density, low cost, and high Li utilization [1,2], it has great potential for automotive applications. The spinel structure offers a 3D network for the conduction of $Li^+$ ions in the solid phase, but in overall, the ionic diffusion coefficient is low, like in most LIB cathode materials. The typical particle size of battery-grade LNMO is in the range of 6-10 μm. Despite the recent interest in single crystal materials initiated by the excellent performance reported for single crystal $LiNi_{0.5}Mn_{0.3}Co_{0.2}O_2$ (NMC532) [3] and $LiNi_{0.5}Mn_{0.3}Co_{0.2}O_2$ (NMC622) [4], current practical materials are polycrystalline. In order to optimize synthesis routes and obtain stable materials with the highest possible ionic conductivity, it is paramount to control the grain boundary structure and chemistry. It is therefore vital to understand the segregation of solutes at grain boundaries. To this end, the present study employs (scanning) transmission electron microscopy (STEM) and atom probe tomography (APT) to investigate the segregation behavior at the grain boundaries of this particular Co-free cathode material in its pristine condition. TEM is an excellent tool for structural analysis down to the atomic scale, but has some shortcomings in terms of chemical detection limits, especially for light elements like Li [5]. In contrast, APT provides three-dimensional compositional mapping at nanometer resolution, and it has a chemical sensitivity in the range of parts per million for all elements, including Li, making this technique ideally suited to complement TEM investigations of LIB cathodes [6].

For instance, APT has been used to investigate Li-loss during cycling and the effect of storage in air in layered nickel manganese cobalt oxides (NMC) and LNMO, often in combination with STEM for structural and chemical (other than Li) analysis [7–9]. The spinels LMO and LNMO have also previously been investigated with APT, where the work by Scipioni *et al.* [10] focuses on determining the chemical composition of the cathode electrolyte interface (CEI) on LMO. APT has also been used to understand different degradation mechanisms in graphite and silicon anode materials [11,12].

Grain boundaries have been the subject of some studies in the past. However, there is still a general lack of available studies on grain boundary segregation that will allow us to better understand the relationship between grain boundary segregation and the performance of the Li-ion battery systems. Recently, a grain boundary investigation was performed on the solid state electrolyte $Li_7La_3Zr_2O_{12}$ (LLZO), to better understand the Li dendrite formation in solids [5]. Maier *et al.* [13] reported the segregation of impurities, such as Na, at the grain boundaries and the appearance of a Ni-rich foreign phase as observed by APT. Currently, there is no published study at the near-atomic scale on the segregation of solutes at the grain boundaries of the Co-free spinel $LiNi_{0.5}Mn_{1.5}O_4$ cathode material, which is the focus of this work.

Specimens for TEM and APT analysis were made from pristine LNMO by using an FEI Helios G4 UX focused ion beam (FIB). To obtain a high-quality surface on the TEM lamella, final thinning was performed with the ion beam set to 2 kV. Scanning precession electron diffraction (SPED) was performed on a Jeol JEM 2100F with a field emission gun (FEG) operated at 200 kV with a Nanomegas Astar precession set-up. The precession angle was 1.0°, and the pixel time was 10 ms, the same time as one full precession period. The resulting SPED patterns were recorded on a Medipix3 MerlinEM direct electron detector with a single 256×256 Si chip from Quantum detectors. The data stacks were processed and analyzed using Pyxem [14], HyperSpy [15], and MTEX [16]. Orientation analysis is based on template matching [17]. The scanning TEM (STEM) analysis was performed on a Jeol JEM ARM 200F equipped with a cold FEG operated at 200 kV. For spectroscopic analyses, a Gatan Quantum ER GIF for dual electron energy-loss spectroscopy (EELS) with dispersion of 0.25 eV/channel to detect all relevant edges simultaneously. A Jeol Centurio SDD (nominal solid angle 0.98 sr) for X-ray energy-dispersive spectroscopy (EDS) was used. The EELS data were analyzed using Gatan Digital Micrograph 3.4.3. The power law was used for background modelling, the Hartree-Slater was used as a cross-section model, and plural scattering was accounted for by utilizing the low-loss spectrum. The EDS data was quantified using the Cliff-Lorimer method [18] in HyperSpy, with calculated *k*-factors taken from Gatan Digital Micrograph. The APT specimens were prepared from site-specific lift-outs following procedures described in detail in [19]. The APT specimens were analyzed using a Cameca LEAP 5000XS instrument operating at 15 pJ laser energy, 250 kHz, and 50 K. The commercial package AP Suite 6.3.1 was used for data reconstruction and analysis. Note that the TEM and APT analyses were not performed in a correlative approach. Instead, different grain boundaries were analyzed by TEM and APT.

A scanning electron micrograph of a representative pristine LNMO particle used in this study is shown in Figure 1a. Figure 1b shows a cross-sectional TEM lamella with a corresponding SPED-based orientation map in the electron beam direction given in Figure 1c. The polycrystalline nature of a single LNMO particle is evident. Based on the investigated area, the grain orientation and the misorientation distribution appeared to be random. For detailed analysis a grain boundary was selected. In Figure 1c, the green grain (inside the dashed black box) is at the [12 2 11] zone axis (ZA), i.e. near a 101 ZA, and the pink grain is at the [7 -6 16] ZA, which gives a misorientation angle between these grains of 36.0°.

The selected green grain was tilted to the [1 1 0] ZA by using Tiltlib [20], as demonstrated by the inset selected area diffraction (SAD) pattern in Figure 1d. This resulted in a near edge-on grain boundary (relative to the electron beam). In this orientation, the grain boundary in the dashed box was studied in detail by EDS and EELS, while an HAADF-STEM image is given in Figure 1e. The high-angle annular dark-field (HAADF) STEM in Figure 1e depicts the spinel crystal structure projected along [1 1 0] above the grain boundary and a lower symmetry ZA below the GB. By utilizing the orientation information obtained by the SPED scan, this orientation is identified as [7 -5 12], which is close to [1 -1 2].

The elemental segregation was studied via EELS and EDS analysis and the corresponding results from this grain boundary are shown in Figure 1f. A distinct segregation of Mn and Na is observed at the grain boundary, while Ni exhibits a depletion at the grain boundary. Similar segregation of Na at grain boundaries in a Li-rich NMC cathode material has also been previously confirmed by TEM [21], but also by APT in lithium manganese oxides [13]. In the same study, the depletion of Mn and the enrichment of Co at grain boundaries were observed. In the current study, Co is removed from the bulk composition, thus this can potentially rationalize the contradicting segregation behavior of Mn. In a different study, no segregation of Mn, Ni, or Co was observed at a grain boundary of monoclinic $Li_2MO_3$, while only after electrochemical cycling a depletion of Mn and enrichment of Ni at grain boundaries was found [22]. It becomes apparent that both the structure and the bulk composition of the cathode material alter the segregation of solutes at grain boundaries. Due to the constraints imposed by the dispersion, which precluded the concurrent verification of Na with the other elements, the fine structure of the Mn_L-edge was not subjected to further analysis. Thus, the Mn oxidation was not further investigated in this study.

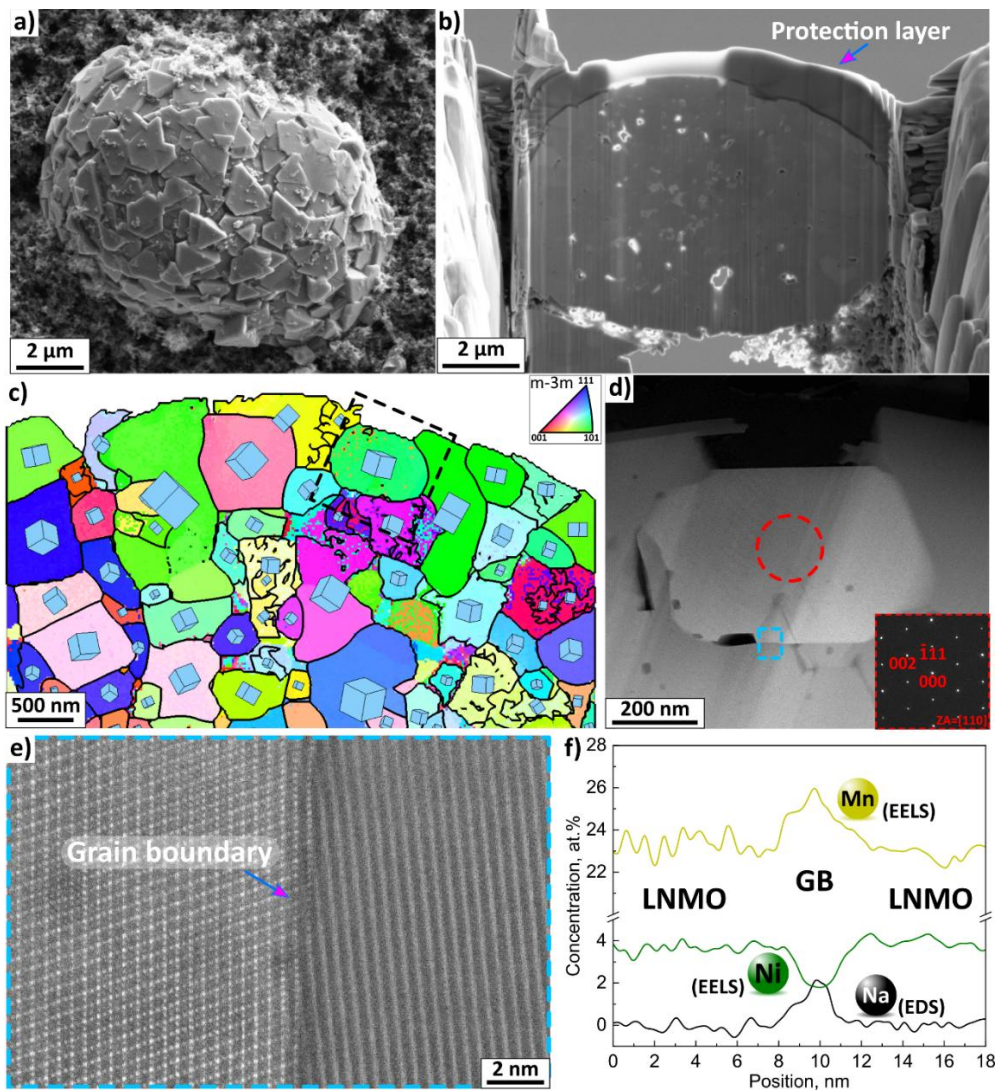

Figure 1: a) Secondary electron micrograph of a pristine LNMO particle. b) TEM lamella prepared by FIB. c) SPED orientation map from the lamella shown in b), where the dashed black box indicates the area of STEM analysis. The inset depicts the coloring for orientations. Model cell orientation is overlayed in addition. d) Medium magnification HAADF-STEM image showing the analyzed area with an inset of selected area diffraction (collected within the dashed red circle) demonstrating that the central grain is tilted to the [1 1 0] ZA. e) Lattice HAADF STEM image of the selected grain boundary indicated by the dashed blue box in d). f) Combined EDS and EELS profiles for Mn, Ni and Na across the area from Figure e).

To further investigate in more detail the segregation of solutes at grain boundaries, particularly those that are challenging for TEM-based or electron-based spectroscopies, APT analysis was also performed for grain boundaries of the pristine LNMO material. Figure 2 shows two APT reconstructions from the pristine material that contain grain boundaries and some Na-rich particles. 1D composition profiles across and perpendicular to the grain boundary in Figure 2a are given in Figure 3. In particular, clear segregation of Mn and O is observed at the grain boundary. Thereby, the segregation of Mn that was measured by EELS in Figure 1 is also confirmed by the APT analysis. At the same time, Li and Ni are depleted from the grain boundary, while Na segregates at the grain boundary in agreement with EDS/ profiles, as shown in Figure 1. In addition, Supplementary Figure 1 shows the atom count profiles for O, Mn and Li corresponding to the 1D profiles from Figure 3a. A clear increase in the number of atoms of Mn and O can be observed at the grain boundary, while the number of atoms for Li decreases. Thus, the observed depletion of Li at grain boundaries does not result in an artificial increase in the composition of the remaining solutes.

Segregation of Na at grain boundaries and Na-rich particles was previously also observed by APT in a $LiMn_2O_4$ material [13]. It is hypothesized that Na originates from the synthesis route of the LNMO, which remains undisclosed in this study due to the use of a commercial LNMO. However, it is often reported in the literature that co-precipitation is used for the synthesis of LNMO, which involves the use of NaOH for the synthesis of the LNMO precursor [23,24], or $Na_2CO_3$ [25], or $Na_2S$ [26]. At the same time, the depletion of Ni at the grain boundaries of the pristine LNMO material is also confirmed by the APT analysis. The type of solute that segregates at the grain boundaries is consistent between the two different regions of interest indicated by the arrows #1 and #2 in Figure 2a. The amount of these solutes in the two regions of interest is similar for all the elements. However, a difference emerges in the extent of nickel depletion, as evidenced by the arrows in Figure 3c and Figure 3d. In particular, Ni is depleted down to 6.0 at.% along the arrow #1 (Figure 3c), whereas along arrow #2 Ni is only depleted down to approximately 6.8 at.% (Figure 3d). A more thorough discussion is provided below, where additional APT data from a separate reconstruction is presented.

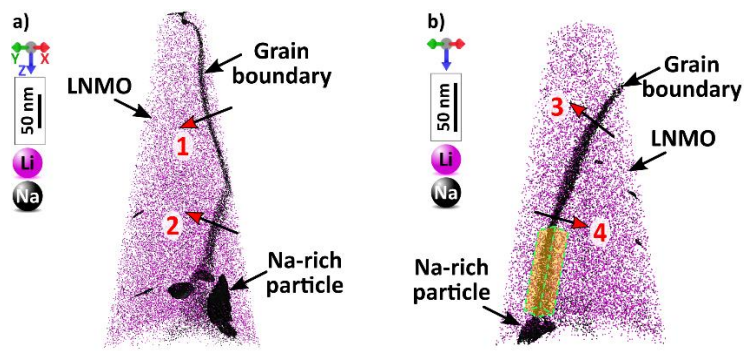

Figure 2: a, b) APT reconstructions from the pristine LiNi$_{0.5}$Mn$_{1.5}$O$_4$ cathode material showing grain boundaries decorated with Na. Na-rich particles are also shown. The arrows indicate the location of the 1D composition profiles shown in Figure 3 and Figure 4. The semi-transparent yellow rectangular box with dimensions of 42×58×30nm denotes the volume used for the 2D maps in Figure 4.

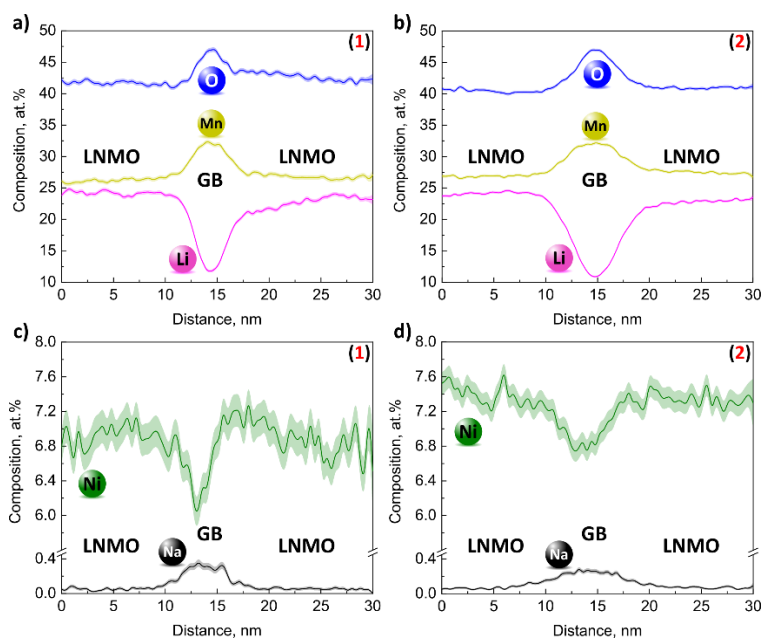

Figure 3: a-d) 1D composition profiles perpendicular to the dislocations as denoted by the arrows #1 and #2 in Figure 2a. Error bars are shown as lines filled with colour and correspond to the 2σ counting error.

To further validate these APT observations, another APT dataset containing a grain boundary was investigated, and it is shown in Figure 2b. Figure 4 shows the 1D composition profiles across and perpendicular to two distinct regions of interest (arrow #3 and #4) at the grain boundary. Similar to all the previous observations (Figure 1f, Figure 3c) and previous studies [13], [21], Na was clearly segregated at the grain boundary. Mn also segregates at the grain boundary, while Li is depleted following a behavior analogous to the profiles in Figure 3. However, in this case, the segregation of O and the depletion of Ni that was previously observed is less significant and clear. In addition, the segregation of Mn and depletion of Li at the grain boundary is less pronounced. Although the absolute values may vary, the general segregation behavior is consistent and comparable between the two techniques. The observed variation in the absolute values can be related to the different types of grain boundaries and misorientations. It is well known that these factors influence the segregation of solutes at grain boundaries [27]. The less pronounced segregation observed may be associated with a grain boundary exhibiting a low-angle misorientation value, while the more pronounced segregation is correlated with a grain boundary characterized by a higher degree of misorientation. Nevertheless, in order to definitively substantiate this correlation, a correlative study is necessary in which either TEM or transmission Kikuchi diffraction (TKD) is conducted prior to the APT analysis.

The necessity for such correlative insights regarding cathode materials is further substantiated by additional observations concerning the segregation of Na. A more thorough examination of the 2D concentration maps of the grain boundary in Figure 2b, corresponding to the semi-transparent yellow rectangular box with dimensions of 42×58×30nm, is presented in Figure 4. The 2D maps demonstrated that the segregation of Na along the grain boundary is discontinuous, as demonstrated in Figure 4e. This finding stands in contrast to the continuous segregation and depletion of Mn (Figure 4f) and Li (Figure 4g), respectively. In a previous correlative TEM/APT study on Si, asymmetric segregation was observed at faceted grain boundaries as a consequence of the local strain state of the facets [28]. While a correlation between the local strain at the grain boundaries and the observed segregation could potentially rationalize this discontinuous segregation, extensive correlative TEM-APT studies are required to unambiguously confirm such a relationship in the LNMO material.

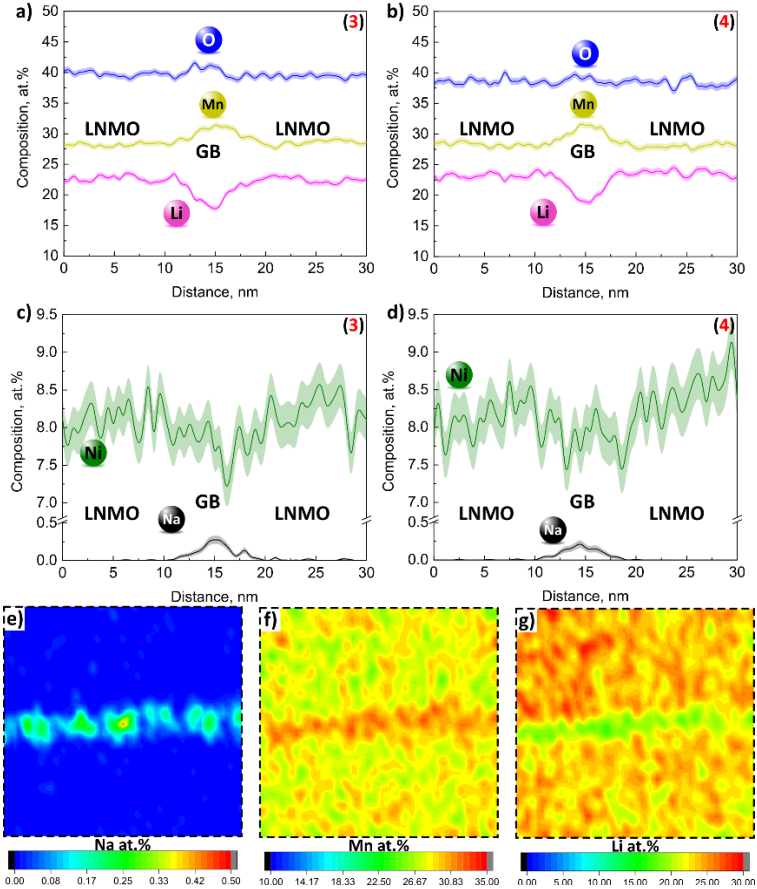

Figure 4: a-d) 1D composition profiles perpendicular to the dislocations as denoted by the arrows #1 and #2 in Figure 2b. Error bars are shown as lines filled with colour and correspond to the 2σ counting error. e-g) 2D concentration maps corresponding to the semi-transparent yellow rectangular box in Figure 2b.

The observed segregation of solutes at grain boundaries can be further rationalized by observations made on solutes that interact with dislocations in the LNMO material. In particular, Figure 5a shows an APT reconstruction from the pristine LNMO material, where tubular features that correspond to elemental segregation at dislocations are observed. Segregation at dislocations is commonly observed by APT in several material systems [29]. For instance, solutes were found to segregate at dislocations and stacking faults in alloys such as nickel-based superalloys and titanium alloys, resulting in microstructural and chemical alterations [30–32]. A 1D composition profile across a dislocation, as denoted by arrow #1 in Figure 5b reveals the segregation of O and Mn at the dislocation, while Li is depleted, as shown in Figure 5f. This segregation is consistent with the observations documented earlier at grain boundaries in Figure 2. In addition, the dislocations in Figure 5b exhibit a structural alignment, suggesting the occurrence of a dislocation pile-up, which may ultimately result in the formation of a low-angle grain boundary.

The observed segregation is further confirmed by 2D compositional maps corresponding to the rectangular region of interest with dimensions of 16×32×10nm in Figure 5b. Figure 5c, d and e show the 2D maps of Li, O and Mn, respectively. It is evident that there has been a marked depletion of Li at the dislocations, accompanied by an enrichment of O and Mn. The enrichment levels exhibited some variation among the three dislocations. It has previously been reported that a similar observation regarding differences in the amount of segregation at the pile-up of dislocations has been made in the case of nickel-based superalloys [33]. Note that Ni does not exhibit any particular segregation interaction with the dislocation.

The formation time of dislocations remains undetermined, given the lack of disclosure regarding the synthesis process. However, a white paper from the manufacturer mentions strain annealing during the sintering process [34]. This process potentially times the formation of crystal defects prior to the sintering process. In the present study, TEM observations revealed the absence of highly strained regions, thereby indicating the overall success of the strain annealing step. This observation suggests that the observed segregation at dislocations is likely an isolated occurrence. Nevertheless, this finding provides fundamental insights that further support the observed grain boundary segregation behaviour. Segregation at dislocations after annealing has been observed in a different material system, where it is hypothesized that the segregation stabilized the fault [35]. Therefore, the annealing process proves inadequate to fully remove the crystal defect. However, this hypothesis requires further investigation at the various stages of the material synthesis process, a topic that extends beyond the scope of the present study.

The uncertainty on the formation time of dislocations creates various scenarios that can potentially explain the lack of Na at the dislocations in this particular case. If dislocations formed after Na has segregated at the grain boundaries, it is apparent that there is no available Na to segregate at these dislocations. However, in Figure 4e, the asymmetric and non-continuous segregation of Na can potentially correspond to segregation at a pile-up of dislocations that form the grain boundary. Figure 5e shows 1D composition profiles across the dislocation denoted with arrow #2. In this case, only enrichment of Mn and depletion of Li at the dislocation is observed. This observation correlates with the grain boundary observations made in Figure 2b. A hypothesis related to the correlation between the type of crystal defect and the segregated solute can rationalize this difference. It is known that different types of dislocations, i.e. screw or edge, will interact differently with the various solutes in a material system. As a consequence, different amounts of solutes will segregate at each individual type of dislocation [36]. Thus, potentially in the current studied LNMO, a network of various types of dislocations exists, leading to disparities in the nature of solute segregation at these dislocations.

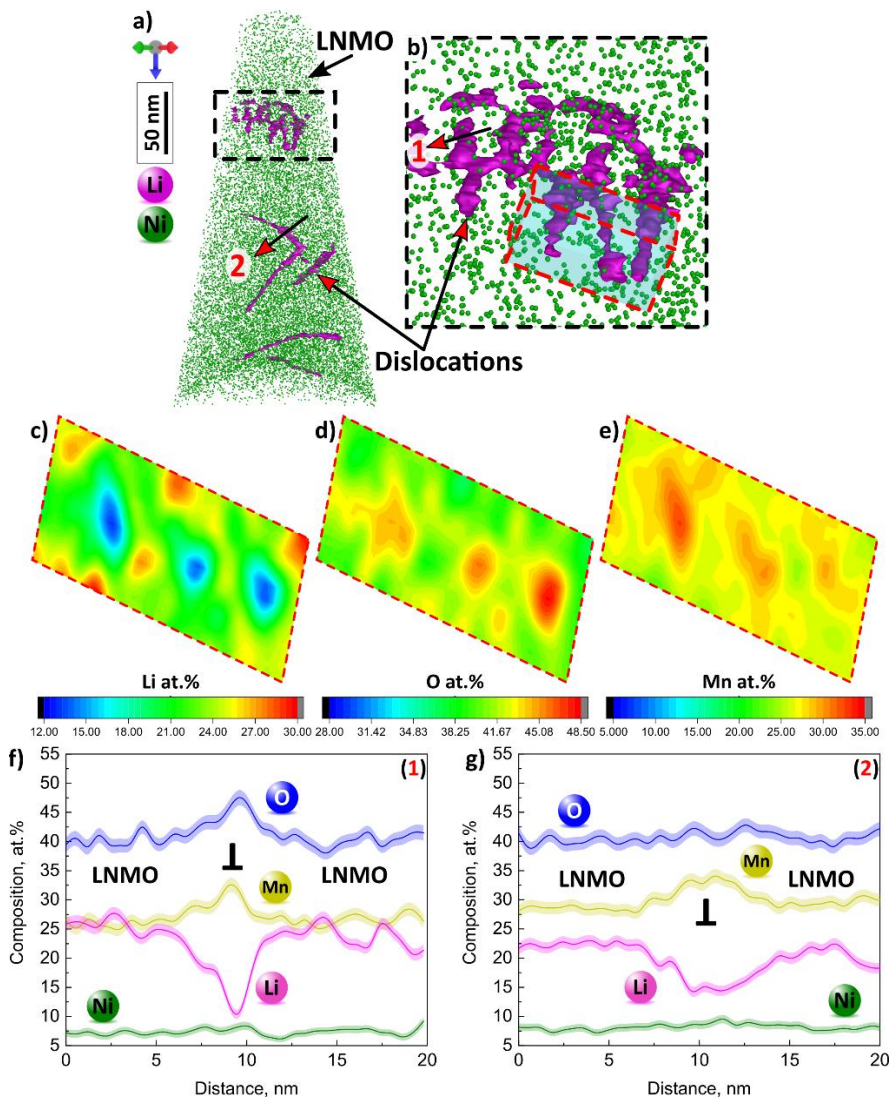

Figure 5: a) APT reconstructions from the pristine LiNi$_{0.5}$Mn$_{1.5}$O$_4$ cathode material showing dislocations within a grain. b) Zoomed-in detail of the dislocations denoted by the dashed black square in a). c, d, e) 2D composition map of Li, O and Mn, respectively, corresponding to the semi-transparent blue rectangular box in b). f, g) 1D composition profiles across dislocations as denoted by arrows #1 and #2, respectively. Error bars are shown as lines filled with colour and correspond to the 2σ counting error.

In summary, the present study investigated the segregation of solutes at grain boundaries in a pristine spinel LiNi$_{0.5}$Mn$_{1.5}$O$_4$ cathode material using transmission electron microscopy and atom probe tomography. The segregation of O, Mn and Na was observed at the grain boundaries. At the same time, Li and Ni exhibit a depletion at the grain boundaries. In certain instances, the segregation of O and depletion of Ni at the grain boundary are not pronounced, while Mn segregation and Li depletion remain evident. The grain boundary segregation is further rationalized by observations made on solutes segregating at dislocations in the pristine LNMO material. In particular, clear segregation of O and Mn and a depletion of Li are observed at dislocations. Besides, in certain instances, the segregation of O is not observed, while the segregation of Mn and the depletion of Li at the dislocation persist. The aforementioned observations in this pristine LNMO material have the potential to establish the foundation for a comprehensive understanding and improvement of its synthesis routes, and eventually its electrochemical performance, thus enabling the green transition.

**Acknowledgments**
The Research Council of Norway (RCN) is acknowledged for its support to the Norwegian Micro- and Nano-Fabrication Facility, NorFab (No. 295864) for the use of the FIB facilities, the Norwegian Laboratory for Mineral and Materials Characterization, MiMaC (No. 269842) for the use of the Atom Probe facility, and lastly to the Norwegian Center for Transmission Electron Microscopy, NORTEM (No. 197405) for the use of the TEM facilities. The final acknowledgment is given to the Norwegian Centre for Environment-friendly Energy Research (FME), co-sponsored by the Research Council of Norway (project number 257653) and 40 partners from research, industry, and the public sector.

**Conflict of interest**
The authors declare that they have no conflict of interest.

# Supplementary information

# Atomic-scale insights on grain boundary segregation in a cathode battery material

Inger-Emma Nylund[a], Elise R. Eilertsen[a], Constantinos A. Hatzoglou[a], Kaja Eggen Aune[b], Ruben Bjørge[b,c], Antonius T. J. van Helvoort[b], Ann Mari Svensson[a], Paraskevas Kontis[a,*],

[a] Department of Materials Science and Engineering, NTNU Norwegian University of Science and Technology, 7491 Trondheim, Norway
[b] Department of Physcis, NTNU Norwegian University of Science and Technology, 7491 Trondheim, Norway
[c] SINTEF Industry, 7465 Trondheim, Norway

*corresponding author:* paraskevas.kontis@ntnu.no

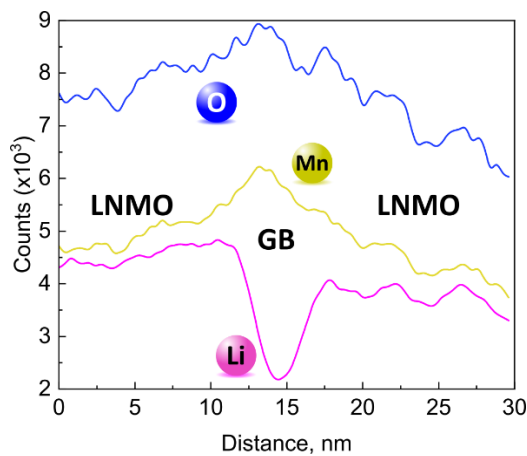

Supplementary Figure 1: Atom count profiles for O, Mn and Li corresponding to the 1D profiles from Figure 3a.